\documentclass[numreferences]{kluwer}    

\newdisplay{guess}{Conjecture}
\usepackage[dvips]{graphicx}

\begin{document}
\begin{article}
\begin{opening}
\title{
Dynamics of a Gear System with  Faults in Meshing Stiffness
}
\author{GRZEGORZ \surname{LITAK}$^1$}
\author{MICHAEL I. \surname{FRISWELL}$^2$}

\institute{~\\$^1$Department of Applied Mechanics, 
Technical University of Lublin, \\ Nabystrzycka 36, 
PL-20-618 Lublin, Poland \\
$^2$Department of Aerospace Engineering, University of  Bristol, Queens
Building, \\ Bristol BS8 1TR, UK}

\date{11 May 2004}

\begin{abstract}
Gear box dynamics is characterised by a periodically changing stiffness.
In real gear systems, a backlash also exists that can lead to a loss in
contact between the teeth. Due to this loss of contact the gear has
piecewise linear stiffness characteristics, and the gears can vibrate
regularly and chaotically. In this paper
we  examine the effect of tooth shape imperfections and defects. Using standard 
methods for nonlinear systems we examine
the dynamics of gear systems with various faults in meshing stiffness.
\end{abstract} 
\keywords{gear system, nonlinear vibrations, meshing errors}
\end{opening}

\section{Introduction}
Gears are very common systems, and practically impossible to replace  
in various applications where mechanical power must be transferred.
Time varying mesh stiffness due to multiple teeth contact 
and a backlash between the teeth give rise to
complex behaviour 
[1--6]. 
In consequence, under a dynamic load, a typical gear system is
a nonlinear oscillator, exhibiting a range of complex behaviour including chaos
[4,7--13].
In use the geometric parameters of the gears change, and this causes 
the corresponding nonlinear response to change 
[4--5,14--15].
Choy {\em et al.} \cite{Cho96} and Kuang and Lin \cite{Kua01} examined the effect of tooth wear.
Vibrations have also been modelled by including stochastic forces  
[4,14--15,18].

In practice it is important to minimise the effect of noise
and keep the machine as close as possible to a stable response. In this paper we 
classify meshing faults and examine the effect of broken teeth and meshing 
stiffness fluctuations on the vibration response. The possiblility of
amplitude jumps in systems with meshing defects is demonstrated.

\section{Modelling of gear dynamics}

Consider the single gear-pair system shown in Fig. 1.
In non-dimensional form, the
equation of motion can be written \cite{War00,Kah97,Lit03} as

\begin{figure}
\begin{center}
\hspace{2.5cm}
\includegraphics[scale=0.6,angle=0]{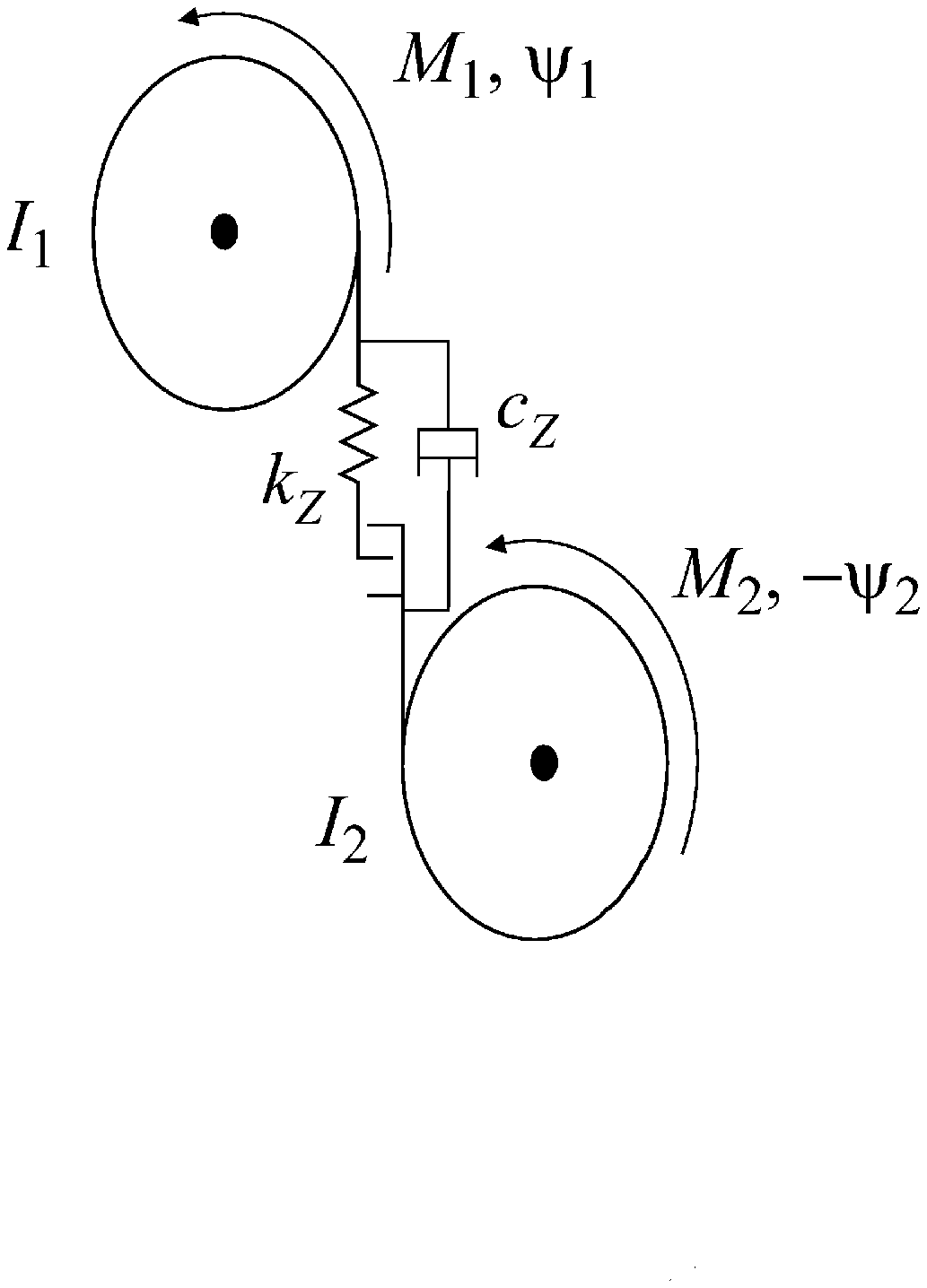}
\vspace{-2cm}
\caption{ \label{fg1} One stage gear system.
}
\end{center}
\end{figure}

\begin{equation}
\frac{{\rm d}^2}{{\rm d} \tau^2}x + \frac{2 \zeta}{\omega} \frac{{\rm d}}{{\rm d} 
\tau}x+ \frac{k(\tau)
g(x,\eta)}{\omega^2}   =  
 \frac{B_0 + B_1 cos (\tau+ \Theta )}{\omega^2},
\label{GLe2}
\end{equation}
where 
\begin{eqnarray}
        \tau &=& \omega t, \nonumber \\
        2 \zeta & = & c_z \left[ r_1^2 /I_1 + r_2^2 / I_2 \right],
 \label{GLe3}
 \\
  B(\tau) & = & r_1 M_1/I_1 + r_2 M_2/I_2. \nonumber
\end{eqnarray}
$\omega$, $\zeta$, $k(\tau)$, $g(x,\eta)$, $\eta$ and $B(\tau)$  
\cite{War00} and 
the other symbols are defined in Table 1 and shown on Fig. 1.

\begin{figure}
\vspace{1cm} \begin{center}
\includegraphics[scale=0.4,angle=-90]{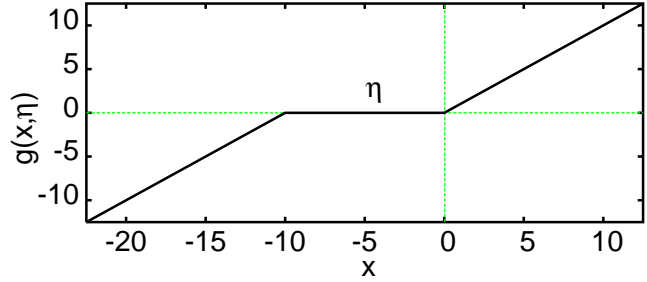}
\caption{\label{fg2} The nonlinear stiffness function $g(x,\eta)$.}
\end{center}
\vspace{0cm}
\end{figure}

\begin{table}
\caption{\label{tableone}
Symbols and parameters used in the analysis.}

\hspace{1cm}
~
\begin{tabular}{||l|l||}
 \hline
$I_1$, $I_2$ & moments of inertia \\
$\psi_1$, $\psi_2$ & rotational angles \\
$x = r_1 \psi_1 - r_2 \psi_2$ & relative displacement \\
$v$ & relative velocity \\
$x_0$, $v_0$ & initial conditions \\
$M_1$, $M_2$ & external torques \\
$\omega$ & excitation frequency \\
$\tau$ & dimensionless time \\
  $\zeta$ &   damping  \\
 $\eta$ &   backlash  \\
  $k(\tau)$ &  meshing stiffness  \\
 $g(x,\eta)$ & nonlinear stiffness function  \\
  $B$, $B_0$, $B_1$ & external excitation  \\
$\delta_i$ &  distance between incerasing teeth contacts \\ 
$\sigma_{\delta}$, $\sigma_k$ & standard deviations \\ 
\hline
\end{tabular}

\end{table}

In the analysis that follows, the stiffness functions $k(\tau)$ and $g(x,\eta)$ need  special 
attention.
$g(x,\eta)$ has a piece-wise character due to the backlash $\eta$, and is shown in Fig. 2. 
$k(\tau)$ is the meshing stiffness arising from the interaction of a single-pair or multiple 
teeth in contact. For an ideal gear system we have followed references \cite{War00,Lit03} and 
assumed that this meshing stiffness changes periodically.
Possible variations from the ideal case, and other possible meshing errors, are plotted in Fig. 3. 

Figure 4 shows the results of simulations of the model given by Eq. (1), with time dependent meshing 
stiffness but without errors.
We have used following system parameters: $\omega=1.5$, $\zeta=0.08$, $B_0=1.0$, $B_1=4.0$, 
$\eta=10$.
With any nonlinear systems multiple solutions may coexist, and the solution obtained depends on the initial conditions. With the above parameters, there are indeed multiple solutions for the gear model, and 
this effect was examined in detail in a previous paper \cite{Sza96}. 
In Figs. 4a and b we show regular and chaotic solutions, depending on the initial conditions
for $[x_0,v_0]=[-9,1]$  and
$[x_0,v_0]=[-9,-1]$, respectively.
Figure 4c shows the two coexisting attractors, obtained for various initial conditions, on the same 
graph.

\section{Errors in meshing stiffness}

\begin{figure}
\begin{center}
\includegraphics[scale=0.50,angle=-90]{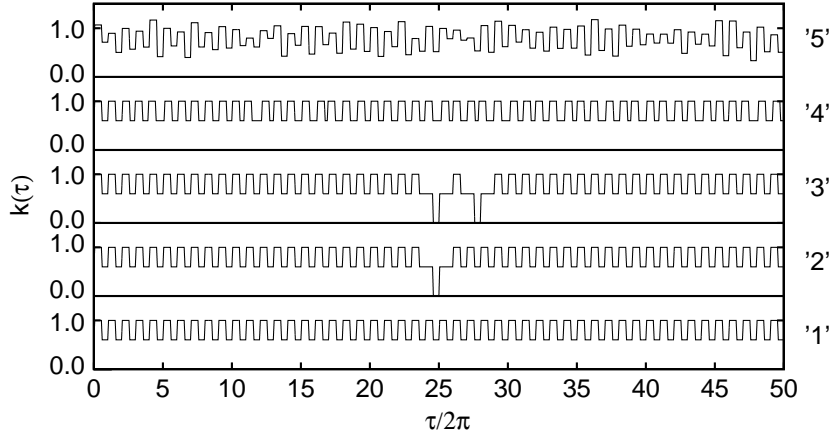}
\end{center}
\caption{ \label{fg3}
Various realizations of meshing stiffness $k (\tau)$. '1'  corresponds to
the ideal system without errors while '2' and '3' show the meshing  
stiffness
with one and two broken teeth. '4' has a randomised distance $\delta_i$  
between increasing teeth contacts  
with a standard deviation $\sigma_{\delta}=0.2{\overline \delta_i}$ 
(${\overline \delta_i}=0.8\pi$ in the non-dimensional time domain), 
and '5' is a randomised  meshing stiffness. Here the amplitude
changes  with  a standard deviation  $\sigma_{k}=0.1$ related to the maximum deterministic value 
$k_{max}=1$.
}
\end{figure}

In this section we examine the effect of meshing stiffness errors.
First consider a gear with 
one or two next neighbour teeth  missing on one of the gear wheels, where
each gear wheel has 50 teeth.
The shape meshing stiffness $k(\tau)$  are given in Fig. 3 as '2' and '3', respectively, 
comparing to the ideal case 
'1'. The response is simulated using the model given in Eq. 1 and modelling the 
meshing stiffness $k(\tau)$ by curves '2' and '3' (Fig. 3). Although the effect of one broken tooth 
appears to be fairly benign in terms of the gears dynamics, if two next neigbour teeth are broken 
the result is a complex response of the system showing the characteristic amplitude jump phenomenon as the solution 
changes from the regular 
to the chaotic attractor. This is visible in Fig. 5, which shows a time history for this case.
Also shown is the reverse jump from the regular to the chaotic attractor, but it is clear 
that the system stays for longer time in the chaotic attractor with intermittent regular motion. This result
 confirms previous results on 
stochastic jumps  \cite{Sat85,War00} in systems with a stochastic force. However, the 
system modelled here is fully deterministic and the broken teeth act as additional parametric excitation.

\begin{figure}

\includegraphics[scale=0.26,angle=-90]{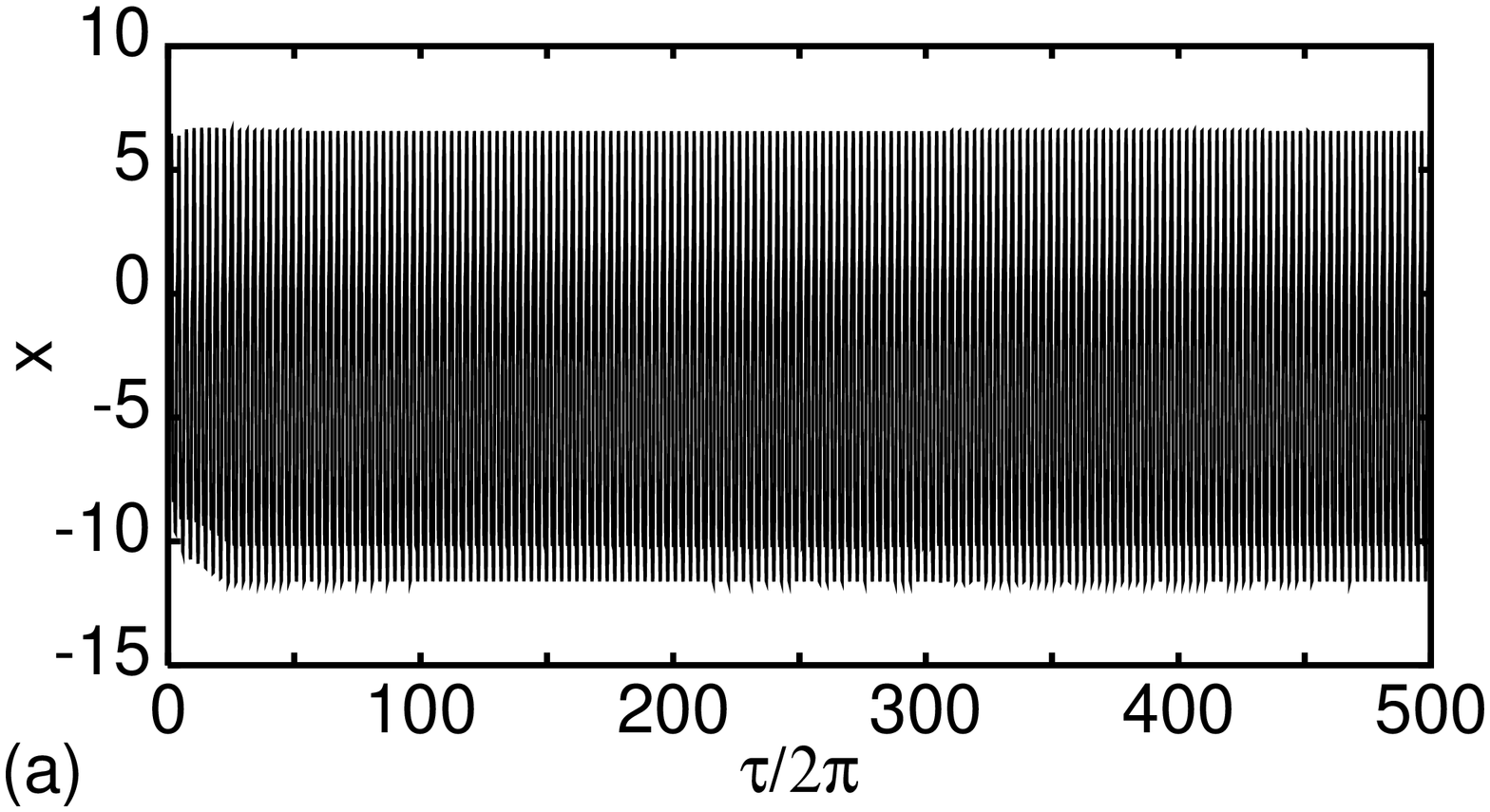} \hspace{-1cm}
\includegraphics[scale=0.26,angle=-90]{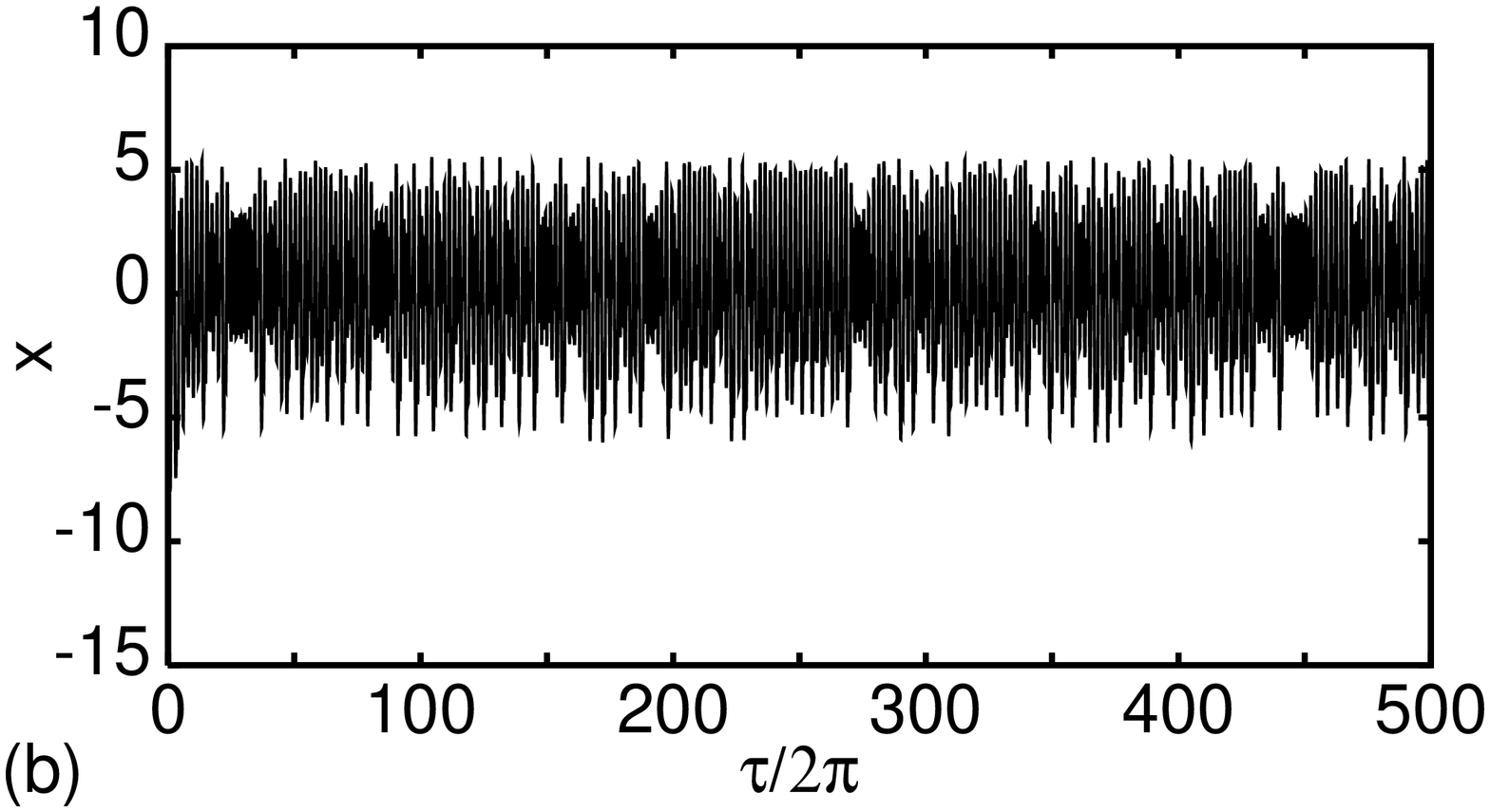}

\begin{center}
\includegraphics[scale=0.4,angle=-90]{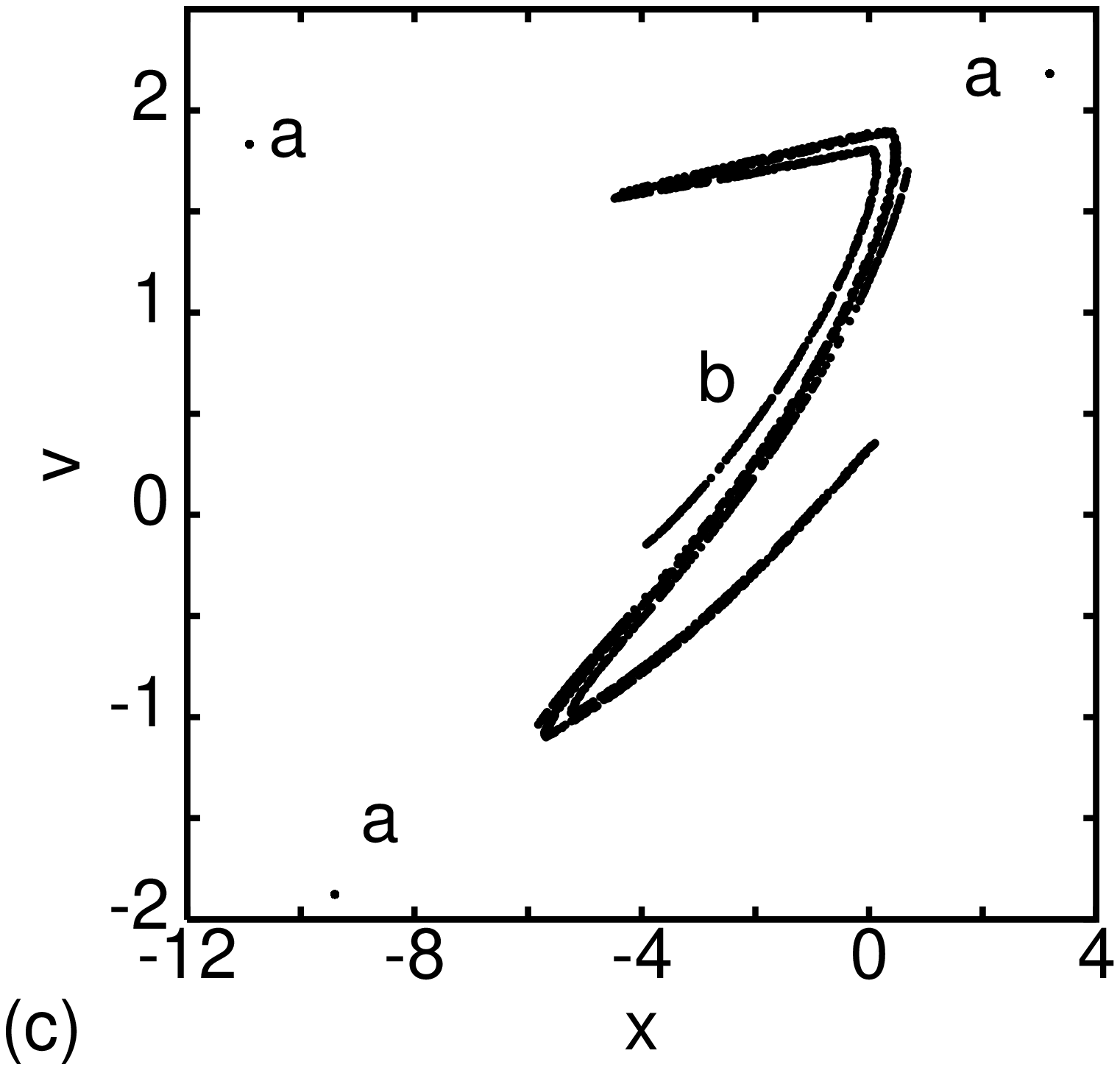}
\end{center}
\caption{\label{fg4}
Time series  and the Poincare map for an ideal system for 
various initial 
conditions $[x_0,v_0]$; regular motion for $[x_0,v_0]=[-9,1]$ (a) and 
chaotic motion for $[x_0,v_0]=[-9,-1]$ (b), respectively. Poincare map for various initial conditions 
(c). The characters a and b in figure c denote the regular and 
chaotic attractors give by time series (a) and (b).  
}
\end{figure}

\begin{figure}
\begin{center}
\includegraphics[scale=0.4,angle=-90]{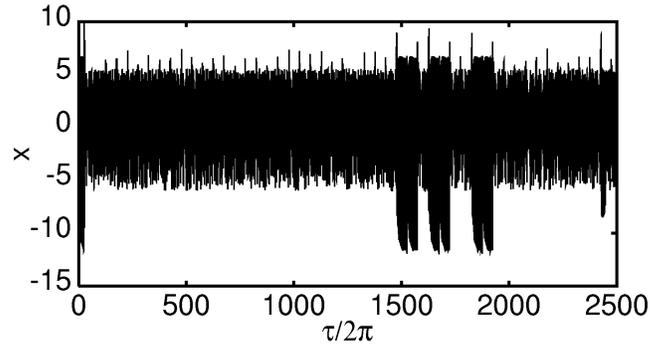}
\end{center}
\caption{ \label{fg5}
Time series for a gear system with two broken neighbouring teeth.
Initial conditions: $[x_0,v_0]=[-9,1]$.
}
\end{figure}

\begin{figure}
\includegraphics[scale=0.26,angle=-90]{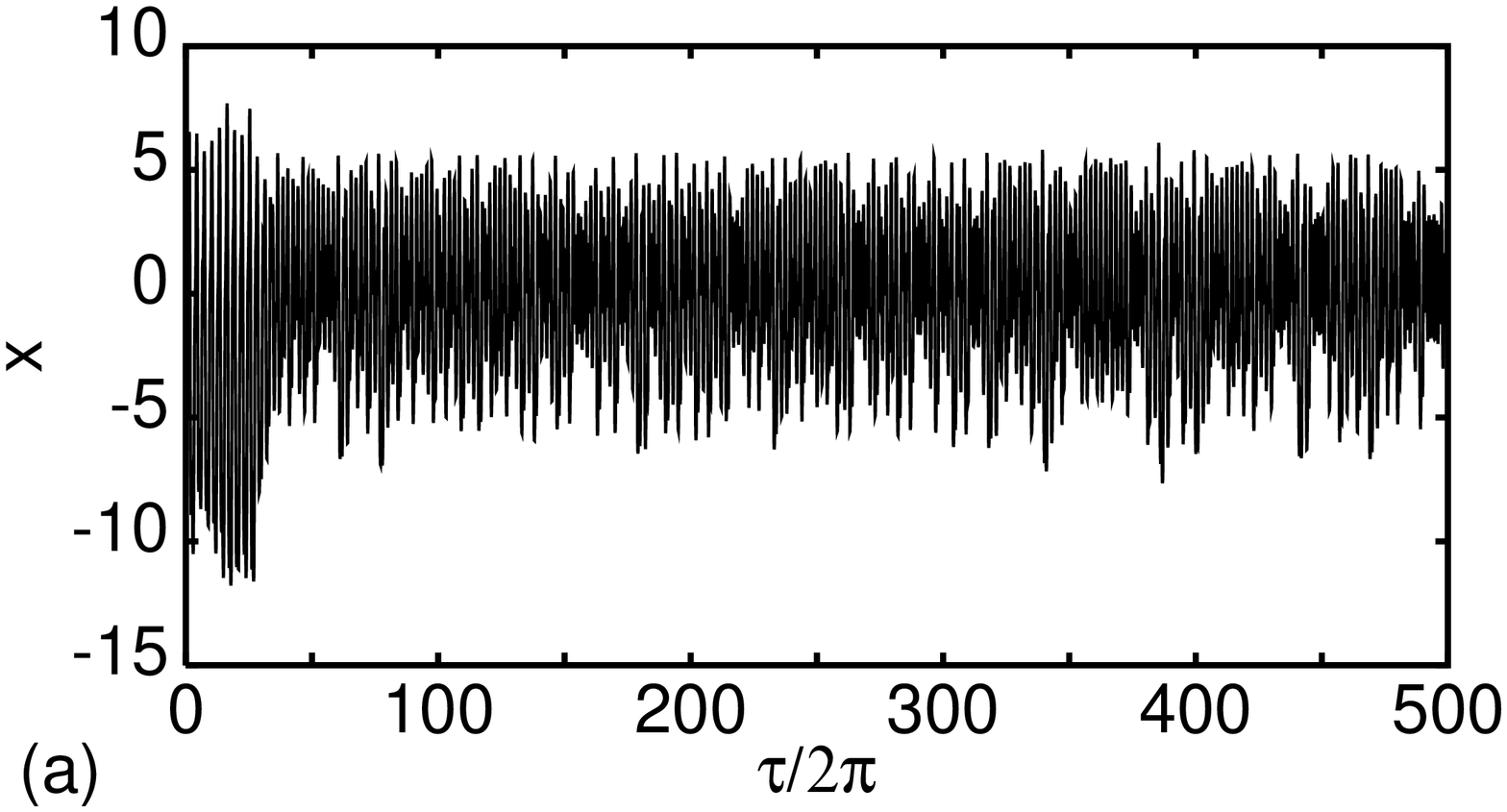} \hspace{-1cm}
\includegraphics[scale=0.26,angle=-90]{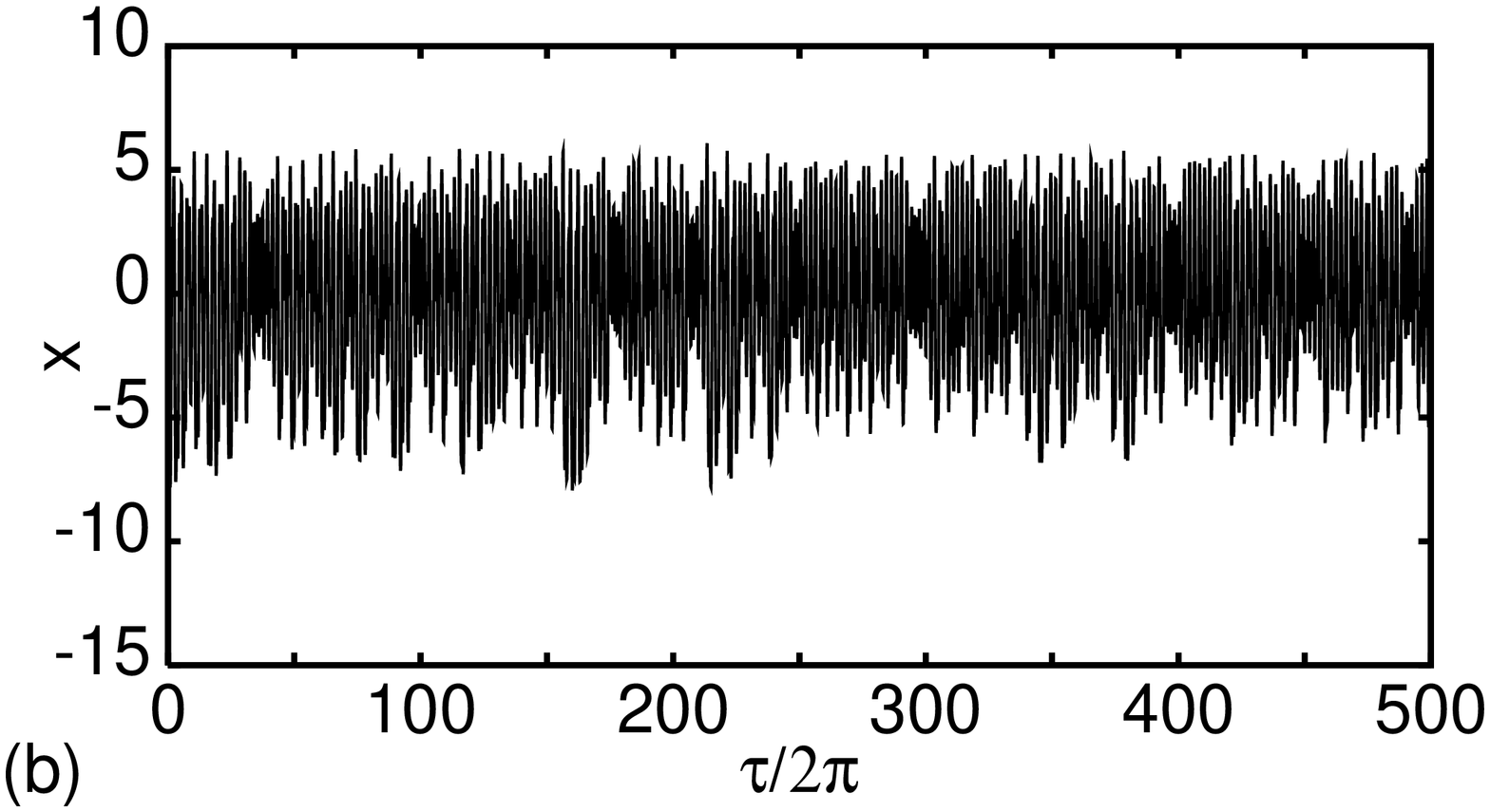}

\includegraphics[scale=0.26,angle=-90]{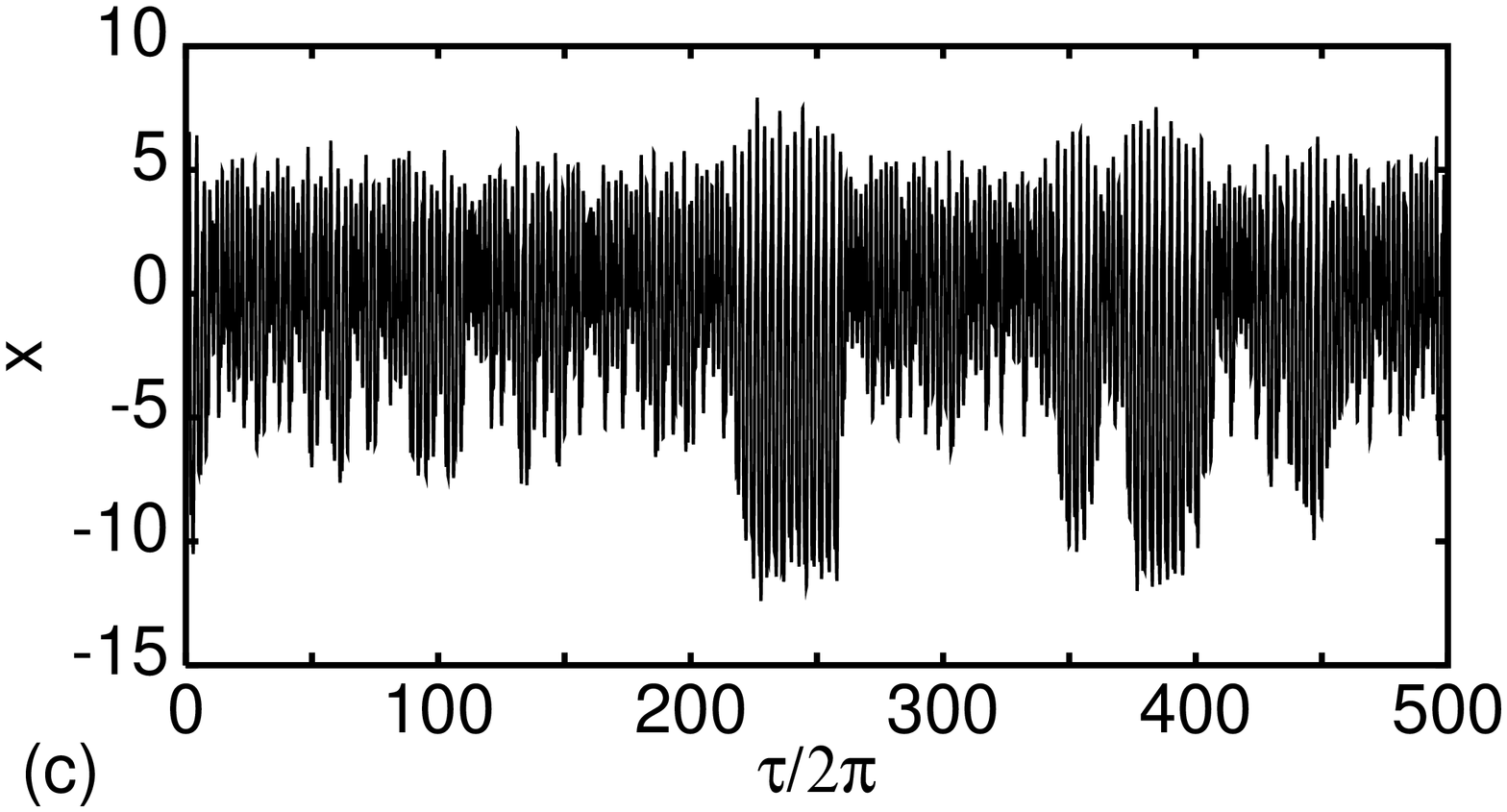} \hspace{-1cm}
\includegraphics[scale=0.26,angle=-90]{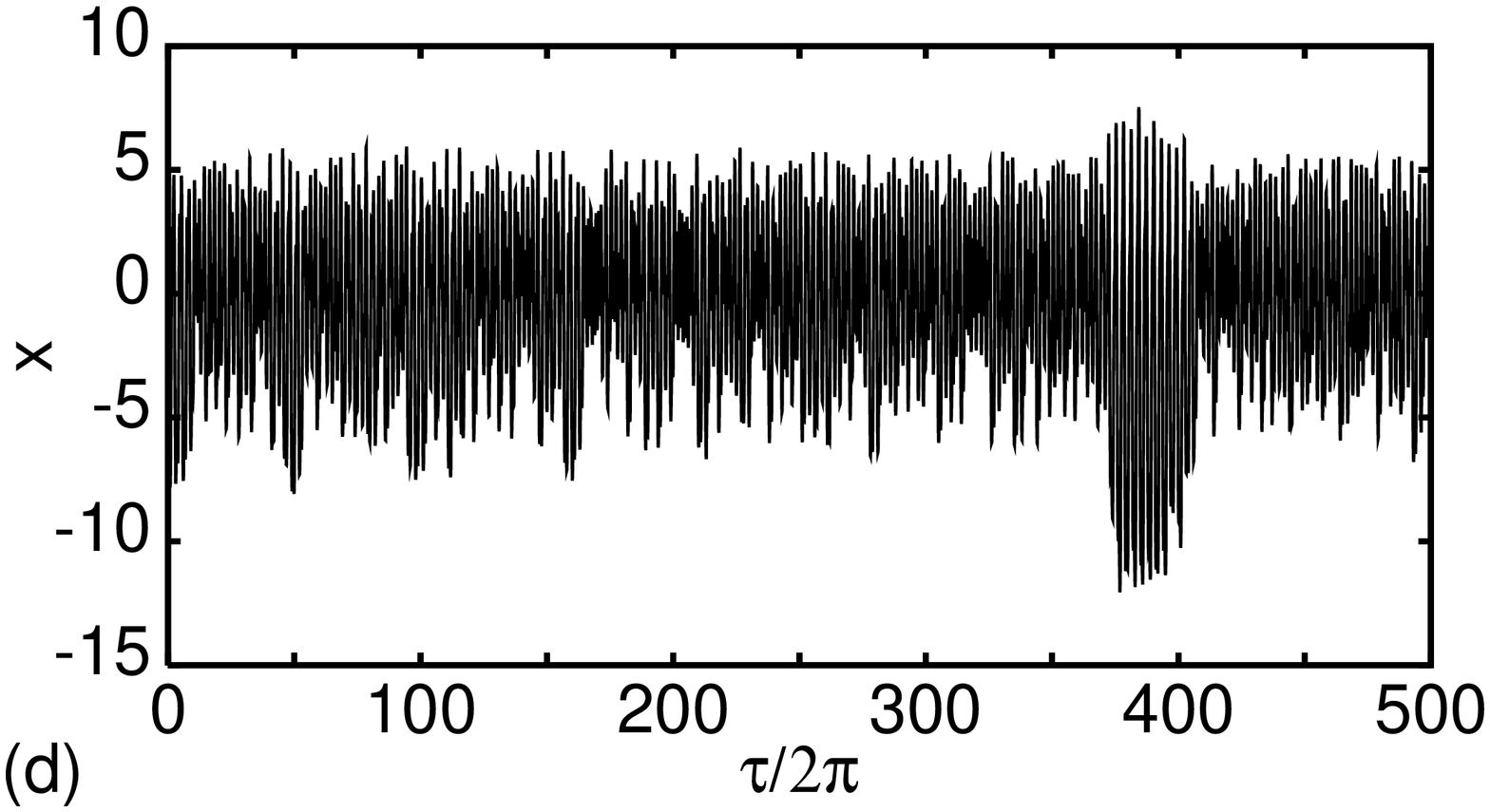}

\caption{ \label{fg6}
Time series  for a gear system 
with a randomised distance $\delta_i$ between increasing teeth contacts
with two  different standard deviations: $\sigma_{\delta}=0.2 \overline{\delta}$ in (a) 
and (b),
and  $\sigma_{\delta}=0.3 \overline{\delta}$ in (c) and (d);
and two different  initial
conditions $[x_0,v_0]=[-9,1]$ in (a) and (c), and
$[x_0,v_0]=[-9,-1]$ in (b) and (d).
}
\end{figure}

\begin{figure}
\includegraphics[scale=0.26,angle=-90]{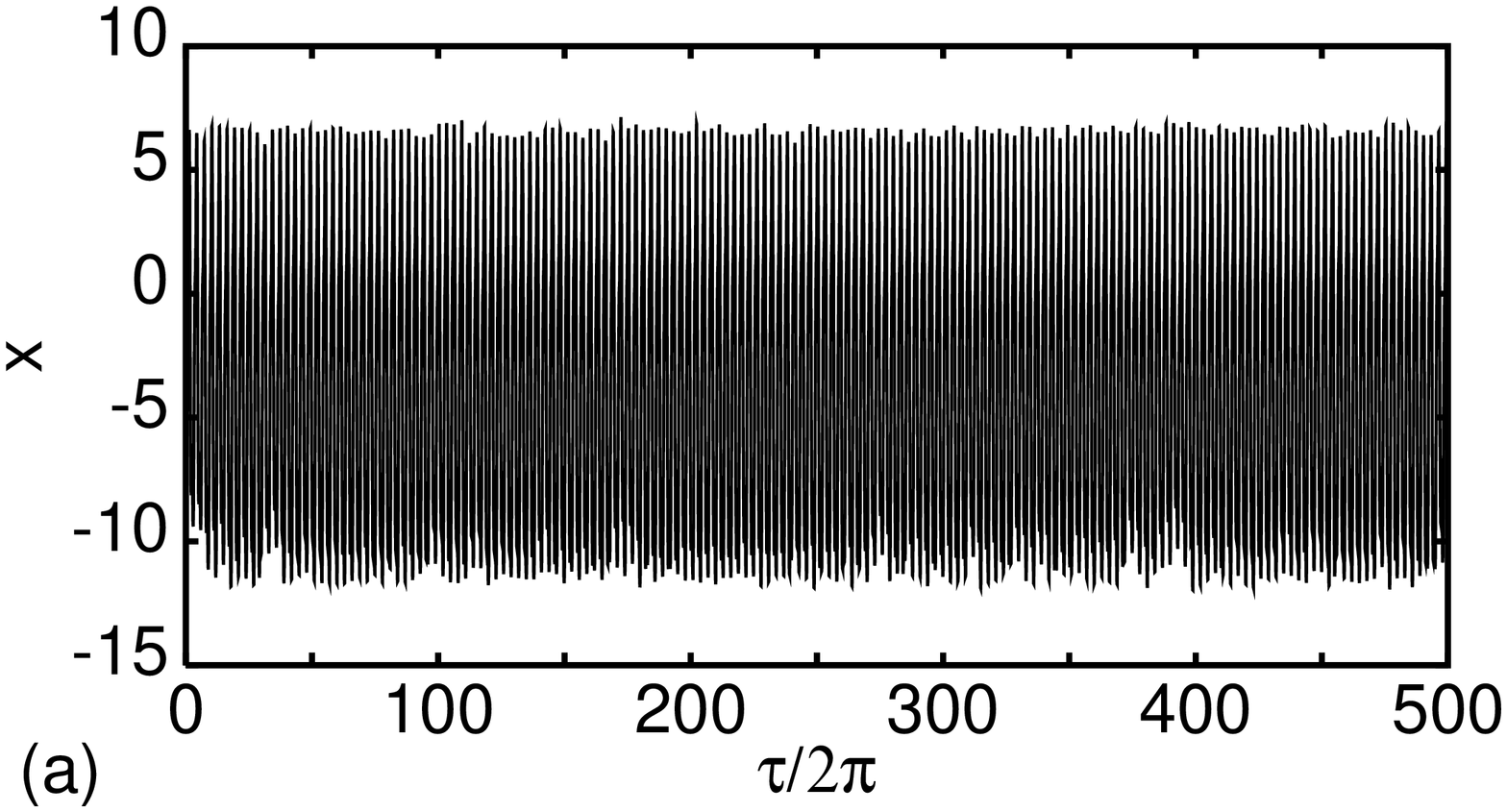} \hspace{-1cm}
\includegraphics[scale=0.26,angle=-90]{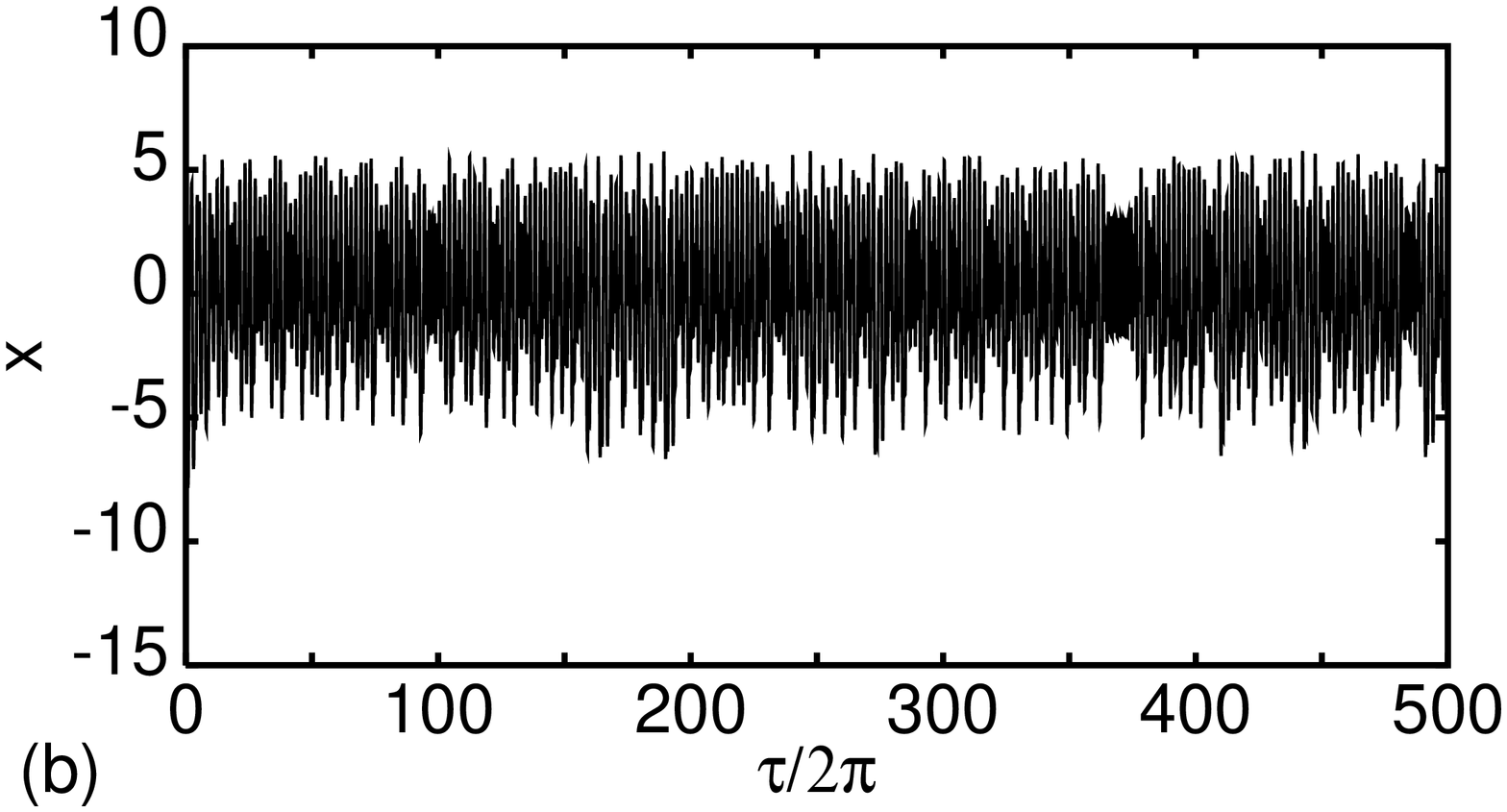}

\caption{  \label{fg7}
Time series for a system with randomised amplitude showing
 regular motion for initial conditions $[x_0,v_0]=[-9,1]$, (a), and
chaotic motion for $[x_0,v_0]=[-9,-1]$, (b).
}
\end{figure}

\begin{figure}
\begin{center}
\includegraphics[scale=0.4,angle=-90]{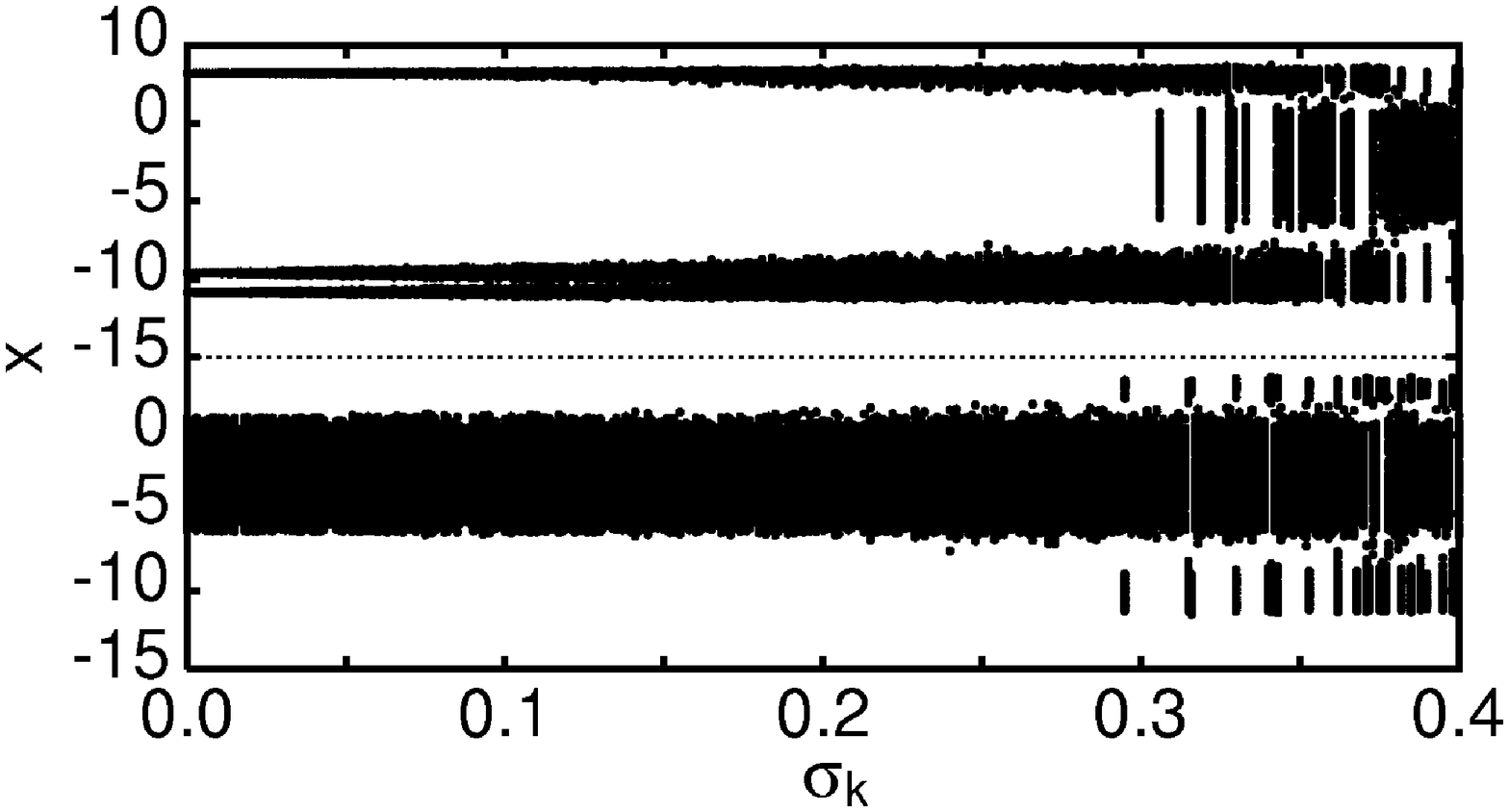}

\end{center}
\caption{   \label{fg8}
Bifurcation diagram for the case of randomised meshing stiffness amplitude: 
$x$ is plotted stroboscopically against the square deviation of fluctuating meshing 
stiffness $\sigma_{k}$ for two different initial conditions
$[x_0,v_0]=[-9,1]$ (the upper panel) and $[x_0,v_0]=[-9,-1]$ (the lower panel).
}
\end{figure}

We have also investigated vibrations of gears
with a random distance $\delta_i$
between their increasing teeth  contacts (Fig. 3, '4'). The results for
two different  noise levels and two different initial conditions $[x_0,v_0]$,
which correspond to different attractors in deterministic case (Fig. 4),
are shown in Figs. 6a-d. Interestingly, for weak noise the system chooses the chaotic attractor.
This conclusion differs from that obtained in the paper by Warmi\'nski {\it et al.} \cite{War00}
but the assumptions about the noise are different.
Warmi\'nski {\it et al.} \cite{War00} used an external stochastic force generated by stochastic Langevin
simulations rather than the stochastic stiffness modelling in the present paper.
For stronger noise the motion shows an intermittent character with short jumps into the regular attractor as
in the previous case with broken teeth (Fig. 5).

Figure 7 corresponds to the meshing stiffness with randomized amplitude but regular distance
(Fig. 3, '5'), for $\sigma_k=0.1$. Here we observe that neither attractor is favored for these
conditions.
Eventually, for a sufficiently large noise level $(\sigma_k \approx 0.3)$
the response returns to the intermittent behaviour with jumps between the regular
and chaotic attractors. This effect is shown clearly in Fig. 8, which shows
 the bifurcation diagram $x$ modulo $2\pi$ against  noise level $\sigma_k$.

\section{Conclusions}

We have examined the dynamics of gears in the presence of meshing faults. 
Such faults may arise due to wear during or incorrect 
tolerance in their production. 
The analysis of 
various types of errors and tooth faults highlights the presence of
dynamic jumping phenomenon. Such jumps between different the types of motions, namely
chaotic and regular, can be crucial for the system reliability.
In this respect our results are consistent with earlier results \cite{Sat85,Lit95,War00}.    
Moreover, the system is more sensitive to errors in the distance between
 teeth than fluctuations in the stiffness magnitude,
although the qualitative effect is similar. One broken tooth has little
influence on the dynamics of thegears, although two broken teeth can have a significant effect.

\begin{acknowledgements}
GL would like to thank the International Centre for Theoretical Physics in  Trieste for
hospitality. MIF acknowledges the support of a Royal Society-Wolfson Research Merit Award.
\end{acknowledgements}

\end{article}
\end{document}